\documentclass[runningheads]{llncs}

\usepackage{graphicx}
\usepackage{amssymb}
\usepackage{pifont}
\usepackage{multirow}
\usepackage[normalem]{ulem}
\useunder{\uline}{\ul}{}
\usepackage{longtable}
\usepackage[hidelinks]{hyperref}
\usepackage{tikz}
\usepackage{booktabs}
\usepackage{graphicx}
\usepackage{adjustbox}
\usepackage{lscape}
\usepackage{caption} 
\usepackage{tablefootnote}
\usepackage{comment}

\begin{document}
\title{KGMM - A Maturity Model for Scholarly Knowledge Graphs based on Intertwined Human-Machine Collaboration}
%
\titlerunning{KGMM - Knowledge Graph Maturity Model}

\author{Hassan Hussein\inst{1}\orcidID{0000-0003-3975-5374} \and
Allard Oelen\inst{1}\orcidID{0000-0001-9924-9153} \and
Oliver Karras\inst{1}\orcidID{0000-0001-5336-6899} \and
Sören Auer \inst{1,2}\orcidID{0000-0002-0698-2864}
}

\institute{TIB Leibniz Information Centre for Science and Technology, Hannover, Germany\\
\email{\{hassan.hussein,allard.oelen,oliver.karras,soeren.auer\}@tib.eu} \and L3S Research Center, Leibniz University of Hannover, Hannover, Germany
}

\maketitle              
\begin{abstract}
Knowledge Graphs (KG) have gained increasing importance in science, business and society in the last years.
However, most knowledge graphs were either extracted or compiled from existing sources. There are only relatively few examples where knowledge graphs were genuinely created by an intertwined human-machine collaboration.
Also, since the quality of data and knowledge graphs is of paramount importance, a number of data quality assessment models have been proposed.
However, they do not take the specific aspects of intertwined human-machine curated knowledge graphs into account.
In this work, we propose a graded maturity model for scholarly knowledge graphs (KGMM), which specifically focuses on aspects related to the joint, evolutionary curation of knowledge graphs for digital libraries.
Our model comprises 5 maturity stages with 20 quality measures. We demonstrate the implementation of our model in a large scale scholarly knowledge graph curation effort.
\end{abstract}

\section{Introduction}

Knowledge graphs have gained increasing importance in science, business, and society in the last years~\cite{Stocker.2022}. However, most knowledge graphs were either extracted or compiled from existing sources. There are few knowledge graphs that were created by an intertwined human-machine collaboration, e.g., Wikidata~\cite{vrandevcic2014wikidata}, Google's knowledge graph~\cite{singhal2012introducing}, or the Open Research Knowledge Graph~\cite{jaradeh_open_2019}.

A number of data quality assessment models have been proposed, since the quality of data and knowledge graphs is of paramount importance~\cite{Bizer2009quality,zaveri2016quality,PipinoData}. A shortcoming of these existing assessment models is that they do not take the specific aspects of intertwined human-machine curated knowledge graphs into account.
Intertwining human and machine collaboration can work in various ways~\cite{oelen2021organizing}. For example, on the one hand, machine intelligence can assist humans in the curation of scholarly knowledge graphs, e.g., by suggesting properties and values to be filled in a curation form. On the other hand, human intelligence can be used to validate the results of machine intelligence, e.g., by voting about the correctness of automatically extracted information pieces. In such scenarios, it is important to assist a community of curators in assessing and \textit{gradually} improving the quality of the scholarly knowledge graph.

   Furthermore, there are some quality problems arise when creating KGs from digital libraries.
   Kroll et al.~\cite{Hermann} ask, ``Why is automatically building knowledge graphs so difficult?" They think that the content selected by digital libraries may be too varied for rule-based methods to produce high-quality knowledge graphs.
   Yadagiri, and Ramesh~\cite{Yadagiri2012SemanticWA} pointed out that organizations' privacy concerns were a significant barrier to linked data technologies adaption in digital libraries. 
   Gonzales~\cite{Gonzales2014LinkingLT} mentioned that copyright restrictions and legalities are the primary barriers for libraries to publishing data on the web. As, the libraries must acquire licenses for numerous periodicals, databases, and other online resources.
   Charles et al. \cite{charles2014links} and Cole et al.~\cite{Cole} remarked that a significant problem for Linked Open Data (LOD) on the web is multilingual and data heterogeneity.
   Sergio et al.~\cite{Oramas2014AutomaticCO} mentioned the challenge of enhancing the user experience when the KG is automatically created from a digital library.
   The ``general validity of facts" is mentioned by Kroll et al.~\cite{Kroll}, as they think that the current knowledge graphs on the web wildly vary from those utilized in scientific digital libraries. Even though entity-centric data in reputable LOD sources on the web may or may not be authentic, it is nevertheless likely to be genuine.
   Cole et al.~\cite{Cole} add that, practical LOD uses in a library setting are still few, immature, and mostly untested.

In this work, we propose a graded knowledge graph maturity model (KGMM), which specifically focuses on aspects related to the joint, evolutionary curation of knowledge graphs.
Our research approach leads to the following research question:
\textit{How can a maturity model for a scholarly knowledge graph look like?}
In that regard, we hypothesize that if the KG developers use our model, then the data will be available for the consumers in the most mature, complete, representable, stable, and linkable shape.

Our model comprises 5 maturity stages with 20 quality measures.
In addition, the measures are prioritized in three categories in each level to further support the applicability of the model.
The model is inspired by the FAIR data principles~\cite{Wilkinson2016}, the Linked Open Data star scheme by Berners-Lee ~\footnote[1]{\url{https://www.w3.org/DesignIssues/LinkedData.html}}, the Linked Data Quality Framework~\cite{zaveri2016quality} but tailors and augments these frameworks specifically for scholarly knowledge graphs intertwining human-machine collaboration. 
Especially for realizing and implementing the FAIR principles aiming at making data \underline{F}indable, \underline{A}ccessible, \underline{I}nteroperable and \underline{R}eusable, we need clear guidance for the developers of knowledge graph applications as well as curators following a principled knowledge graph quality model.

We demonstrate the implementation of our model in a large scale scholarly knowledge graph curation effort with more than 500 collaborators. We show how the model is implemented and can be used to incrementally assess and improve specific parts of the scholarly knowledge graph. Curators are given clear guidelines on how the maturity of a certain part of the scholarly knowledge graph can be improved to reach the next maturity stage.

The article is structured as follows: 
We review related work in Section~\ref{sec:related}. 
The research method is explained in Section~\ref{sec:research-method}.
Our proposed maturity model is introduced in Section~\ref{sec:kg-maturity-model}. 
In Section~\ref{sec:use-case}, we present a use case based on the previously introduced model. 
Finally, we discuss and conclude our work in Section~\ref{sec:conclusions}.

\section{Related Work}
\label{sec:related}
According to Proenca~\cite{proenca_methods_2016}, a maturity model is a methodology used to effectively assess elements of objects, such as processes or organizations, in various domains. Based on this assessment, maturity models enable to evaluate the current maturity level of a respective object in order to contribute to its continuous improvement by providing guidance on how to reach the next maturity level. In a literature review~\cite{orkg_comparison_maturity_models}, we found eleven maturity models  from six different domains, such as Software (Engineering), Information Management, and Business Information, that have an average of five maturity levels (minimum: 4, median: 5, maximum: 6) and cover an average of six attributes (minimum: 3, median: 6, maximum: 30). Table~\ref{table:maturity-comparison} shows an excerpt of this comparison. One of the most well-known maturity models is the Capability Maturity Model (CMM) by Paule et al.~\cite{CMMI-Paulk}. CMM enables software development organizations to enhance their consistency and capability to deliver high-quality software within budget and on time. In this work, we follow the same line of thought but focus on maturity in general for all sorts of knowledge graphs. Similar to CMM, our proposed KGMM enables the assessment of the quality of a knowledge graph in order to determine its current maturity level and thus provide guidance on how to achieve its next level.

A knowledge graph is a special kind of database that manages data in a graph structure. Therefore, the quality of knowledge graphs is closely related to the quality of its data. For any information system, data quality is crucial, as data is the cornerstone of the system~\cite{Larryquality}. Wang and Diane~\cite{Wang1996quality} define data quality as data that is ``fit-for-use" for a certain application or use case so that it fulfills its users' needs. This consideration of data quality is in line with ISO 19113 (2002) that describes data quality as the ``totality of characteristics of a product that bear on its ability to satisfy stated and implied needs''. Consequently, high data quality ensures that a system or product provides benefits for its users by satisfying their needs. In contrast, low data quality has several negative implications for the users, such as decreased customer satisfaction, increased operating costs, ineffective decision-making processes, and poor performance~\cite{PipinoData}. Therefore, any issues in the quality of data harms its use and thus restricts the value of the system or product from the users' point of view~\cite{zaveri2016quality}. 

For this reason, Berners-Lee introduced the star rating system and the Linked Data Principles as instruments for a ``good linked data". This work was the starting point for the development of several approaches that examine and support the quality of knowledge graphs. One of the most well-known and currently prominent approaches is the FAIR principles by Wilkinson et al.~\cite{Wilkinson2016} to boost machines' ability to automatically find and use data (i.e., machine-actionability). Zaveri et al.~\cite{zaveri2016quality} analyzed 30 of these existing approaches, resulting in 23 data quality measures, which they summarized in the Linked Data Quality Framework. This work is one of the most recent and comprehensive frameworks on data quality. We built our KGMM model adopting the work of Berners-Lee, Wilkinson et al.~\cite{Wilkinson2016}, and Zaveri et al.~\cite{zaveri2016quality}, as their work is widely adopted in the data management domain and provides comprehensive guidelines for data quality assurance.

A shortcoming of these works is that they do not consider specific aspects of intertwined human-machine collaboration. However, Zogaj and Bretschneider~\cite{zogaj_analyzing_2014} explain that successful information system projects often also include crowdsourcing. In this context, Wang et al.~\cite{wang_research_2019} further stated that expert crowd members produce extremely accurate knowledge graphs. However, crowdsourcing is often not sustainable due to the limited human resources available. Therefore, crowdsourcing is usually used to refine knowledge graphs for three reasons. First, automatic methods struggle to achieve both high accuracy and broad coverage. Second, the network's documents have a long tail effect, implying that a large amount of knowledge is poorly dispersed. Finally, automatic processing technology is prone to flaws such as excessive noise levels and problems ensuring knowledge accuracy~\cite{qi_knowledge_2021}. We consider these reasons and their effects on the data quality by reflecting these human-machine dimensions in KGMM.

\section{Research Method}
\label{sec:research-method}
For developing and realizing the KGMM we followed a design science approach including the following five-step methodology:

\begin{enumerate}
    \item Reviewing different existing data quality and maturity models~\cite{orkg_comparison_maturity_models}.

    \begin{table}[t]
    \vspace{-2ex}
    \caption{Overview of selected maturity models in various domains.}
    \label{table:maturity-comparison}
    \resizebox{\columnwidth}{!}{%
    \begin{tabular}{|p{3.8cm}|c|p{3cm}|c|p{2.5cm}|}
    \toprule
    \textbf{Module Name} & \textbf{Levels/}  & \textbf{Domain} & \textbf{Maturity} & \textbf{Practicality} \\
    & \textbf{Attributes} & & \textbf{Definition} & \\ \midrule
    Capability Model \newline Integration (CMMI)\tablefootnote[2]{\url{https://resources.sei.cmu.edu/library/asset-view.cfm?assetid=9661}} & 5/ 22 & Software & \checkmark & Specific improvement activities \\ \hline
    Model-driven Development (MDD) Maturity Model~\cite{rios_mdd_2006} & 5/ 3 & Product Lifecycle & - & General \newline recommendations \\ \hline
    Assessing Business-IT~\cite{Jerry_Assessing_Business_IT} \newline Alignment Maturity & 5/ 5 & Business Information Technology & - & General \newline recommendations \\ \hline
    Gartner Enterprise Information Management Maturity Model~\cite{Newman2008GartnerIT} & 6/ 4 & Information Management & - & General \newline recommendations  \\ \hline
    A Capability Maturity Model for Research Data Mgmt.~\cite{qin_capability_2014} & 5/ 6& Research Data \newline Management & - & General \newline recommendations\\ \bottomrule
    \end{tabular}%
}
\end{table}
    \item Eliciting requirements for a KG maturity model from the large-scale knowledge graph application Open Research Knowledge Graph (ORKG) aiming at intertwining human and machine collaboration.
    \item Studying the different quality dimensions in the previous works.
    \item Develop the KGMM by prioritizing and weighting various quality dimensions and measures based on the reviewed literature.
    \item Extension of the ORKG application by implementing features catering for the various KGMM measures and their evaluation. 
    
\end{enumerate}

\section{Knowledge Graph Maturity Model}
\label{sec:kg-maturity-model}
In this study, we are proposing a tool to assess the KG maturity by measuring the data quality at each level.
In the KGMM maturity model, we define and prioritize quality measures for each level to boost the data quality and elevate the KG maturity.
We adapted the three priority levels from the FAIR Data Maturity Model Working Group ~\cite{fair_data_maturity_model_working_group_2020_3909563} and apply them to prioritize the quality measures for each level.
The three priorities are defined as follows:
\begin{itemize}
  \item \textbf{Essential}: denotes a quality metric that is crucial for approaching a maturity level, or, on the contrary, that maturity would be practically impossible to reach if the priority was not fulfilled.
  \item \textbf{Important}: addresses a quality metric that may or may not be vital in some situations. Achieving it, if at all possible, would significantly boost maturity.
  \item \textbf{Useful}: is a type of measure that concentrates on a nice-to-have quality metric that is not required.
\end{itemize}
We also use the pass-or-fail approach from the same publication, the FAIR Data Maturity Model Working Group~\cite{fair_data_maturity_model_working_group_2020_3909563}, to identify whether the quality measures for KG are meeting a given maturity level requirement or not. First of all, we associate each quality metric with a priority (Essential, Important, or Useful).
In any event, if a specific quality metric is ``essential", then the KG must fulfill this quality metric (pass) to advance to the next maturity level.
Furthermore, at least 50\% of the quality measures that are ``important" at a given maturity level should ``pass". Otherwise, the maturity level will be ``fail".
In conclusion, to make the KG pass a certain level, the KG should fulfill all the ``essential" quality measures and at least 50\% of the ``important" quality measures.

In \autoref{table:maturity-levels} we list the seven data quality dimensions in the rows and the five maturity levels in the columns. The 20 quality measures themselves are displayed in the table cells. 

\begin{landscape}
\begin{table}[t]
\vspace{-2ex}
\centering
\caption{\textit{Overview on the KGMM maturity model.} The five maturity levels are listed in the columns and the quality dimensions in the rows. We also indicate priority levels and relevance for human-machine curation. For more concrete examples of the quality problems and the relevance for human-machine curation, please check our detailed table available on Zenodo. \cite{hussein_hassan_2022_6732786}} 
\label{table:maturity-levels}
\resizebox{\columnwidth}{!}{%
\begin{tabular}{|@{}p{2.7cm}|p{3.9cm}|p{4.5cm}|p{3.4cm}|p{3.4cm}|p{3.4cm}@{}|}
\toprule
\textbf{Quality}\newline \textbf{Dimensions} & \textbf{Level 1: \newline Published} & \textbf{Level 2: \newline Completeness} & \textbf{Level 3: \newline Representation} & \textbf{Level 4: \newline Stability} & \textbf{Level 5:\newline Linkability} \\ \midrule
\textbf{Accuracy} & Syntactic accuracy**[H,M] & Timeliness***[H] \vspace{3pt}\hrule\vspace{3pt} Correctness***[H,M]\vspace{3pt}\hrule\vspace{3pt}  Semantic accuracy**[H,M] &  &  &  \\ \hline
\textbf{Completeness} &  & Trustworthiness[H] *** \vspace{3pt}\hrule\vspace{3pt} Instance completeness**[H,M] \vspace{3pt}\hrule\vspace{3pt} Property completeness**[H,M] \vspace{3pt}\hrule\vspace{3pt} Population completeness**[M] &  &  & Linkability*[H,M] \\ \hline
\textbf{Findability} &  &  &  & Identifier stability**[M] &  \\ \hline
\textbf{Accessibility} & Responsiveness***[H,M] \vspace{3pt}\hrule\vspace{3pt} Easiness**[H] & &  & Queryability**[H,M] & Dereferencability *[M] \\ \hline
\textbf{Interoperability} &  & Provenance***[H] & Data representation*[H] & Trackability***[H,M] &  \\ \hline
\textbf{Reusability} & License***[H,M] &  & Reusability***[M] &  &  \\ \hline
\textbf{Succinctness} &  &  & Conciseness***[H,M] &  &  \\ \bottomrule
\end{tabular}
}
{\vspace{2pt}\raggedright*** Essential, ** Important, * Useful\par}
{\vspace{2pt}\raggedright[H] Indicates the relevance to human curation.\par}
{\vspace{2pt}\raggedright[M] Indicates the relevance to machine-actionability.\par}
\end{table}
\end{landscape}

\subsection{Level 1: Published}
At this level, the KG should be published on the web (in any format) with an open license fulfilling the following requirements: 
\begin{itemize}
  \item \textbf{Responsiveness (Essential)}: It is marked as ``essential" because the KG is not useful for the end user if the KG is not accessible in a reasonable time. Nah~\cite{nah_study_2004} recommends that the loading time should be below two seconds.
  The exploration and curation Web interface for the KG should be accessible in a responsive manner with small page loading times.
  
  \item \textbf {Licences (Essential)}: The KG should be licensed under an open licence in order to make the data available for everyone. The licensing information should integrated into the KG interface in a machine-readable manner. The Creative Commons organization created an ontology to provide a machine-actionable representation of license information~\cite{10.2307/j.ctt5vjsx3.16}.
  \item \textbf {Syntactic Accuracy (Important)}: 
  
  The syntactic accuracy according to Hogan et al.~\cite{hogan_knowledge_2022}, is the degree to which the data are accurate in terms of the grammatical rules set for the domain and/or data model. 
  The syntactic accuracy is marked as ``important" because even without complete syntactic accuracy the KG might still be usable for both human and machine.
  Syntactic accuracy can be ensured by restricting the user interface (e.g., by form-based interactions) to only allow the entering of syntactically accurate representations.
  
  \item \textbf {Easiness (Important)}: We mean by easiness, an intuitive way to navigate, explore and curate a KG. 
  Also, Garett et al.~\cite{garett_chiu_zhang_young_2016} find the easiness (easy navigation) as one of the unique (important) design factors that affect user engagement. 
  Easiness also according to Hogan et al.~\cite{hogan_empirical_2012} means, the simplicity with which a user can interpret data without confusion, which includes at least the availability of human-readable labels and descriptions that allow them to grasp what is being presented. 
  
  We mark this measure as ``important" because it is a key factor for sustaining user engagement with a KG.
  
  While the other measures in this layer can be largely automatically validated, checking for easiness requires more qualitative assessment means such as user studies, Web analytic, surveys or interviews.
\end{itemize}

\subsection{Level 2: Completeness}
Wang and Diane~\cite{Wang1996quality} define completeness as the degree to which data has enough breadth, depth, and scope for the task at hand. 
Furthermore, according to Hogan et al.~\cite{hogan_knowledge_2022}, data completeness is the extent to which a data set provides all the needed data. Issa et al.~\cite{issa_knowledge_2021} conducted a review that revealed nine of the most frequently used techniques (automatically or semi-automatically) for assessing dataset completeness.
We adopt semi-automatic techniques to let humans (crowd-members) and machines work jointly to complete KG data.
In Section~\ref{sec:use-case}, we discuss in more detail how we implemented a mechanism that allows intertwined human-machine collaboration.
Based on the previous listed work, we consider complete and current data as a key aspect of KGs. 
Therefore, we define the following quality measures relates to completeness:

\begin{itemize}
  \item \textbf {Correctness (Essential)}: Hogan et al.~\cite{hogan_knowledge_2022} use the term ``validity" to refer to correctness. 
  This is accomplished by imposing restrictions to prevent validity infringements within the KG. 
  According to Batini and Scannapieco~\cite{Batinidata2006}, corrections are essential when the KG data comes from sources that are subject to error (e.g., caused by manual data entry) or from sources whose dependability is unknown.
  We consider correctness as “essential” because incorrect KG data leads to misleading forecasts for the data consumers. If the data is inaccurate, resources, time, and money are wasted.
  \item \textbf {Timeliness (Essential)}: 
  According to Pipino et al.,~\cite{PipinoData} timeliness refers to how current the data is regarding the use case or objective.
  Fürber and Hepp~\cite{Frber2011SwiqaA} think that we can compare two timestamps to discover possible data obsoleteness if at least timestamps describing the last update of data are available in the source and destination data sources.
  We consider timeliness as ``essential" because outdated knowledge can lead to inaccurate analyses, which in turn leads to wrong decisions. Additionally, outdated knowledge can be considered incorrect, which relates to the previously described metric. 
  
  \item \textbf {Provenance (Essential)}: 
  Provenance data is widely acknowledged as essential for facilitating the reuse, management, and reproducibility of published data~\cite{Simmhan05asurvey}.
  According to Golbeck and Mannes~\cite{Golbeck2006UsingTA}, when it comes to filtering and aggregation, tracking the provenance of Semantic Web metadata is essential, especially when the trustworthiness of the data is in question.
  As previously mentioned, we leverage a crowdsourced approach for KG data creation and curation. To ensure data validity, provenance tracking is a crucial aspect to keep track of the history, and to be able to link data to the original source. Therefore, we consider provenance as ``essential". 
  \item \textbf {Trustworthiness (Essential)}: 
   According to Zaveri et al.~\cite{zaveri2016quality}, trustworthiness is the degree to which the data is considered correct, verifiable, actual, and believable. The data trustworthiness, according to Jacobi et al.~\cite{bassiliades_rule-based_2011}, is either: trustworthiness of the source, relative content trust (i.e., depending on the domain knowledge of the source, claims are either trustworthy or not), and a combination of factors. 
    We have marked trustworthiness as ``essential" because when a KG cannot be trusted, the data consumer cannot make a well-informed decision.
  \item \textbf {Semantic Accuracy (Important)}:
  Hidalgo-Delgado et al.~\cite{Hidalgo-Delgado_quality_2021} define semantic accuracy as the level to which data values accurately represent real-world phenomena, as impacted by inaccurate extraction outputs, an imperfect number of claims, destruction, and other considerations.
  \item \textbf{Instance Completeness (Important)}: According to Micic et al.~\cite{Micic2017TowardsAD}, an instance is complete if a dataset includes all of the real-world objects required for a given task.
  \item \textbf{Property Completeness (Important)}:
  According to Batini and Scannapieco~\cite{Batinidata2006}, the property completeness is an estimate of the incomplete extent for any property in a KG.
  Hogan et al.~\cite{hogan_knowledge_2022} define property completeness as the measure of missing values for a given property in a KG.
\item \textbf{Population Completeness (Important)}:
  According to Batini and Scannapieco~\cite{Batinidata2006}, population completeness compares missing values to a reference population. 
  Pipino et al~\cite{PipinoData}. define population completeness as the ratio of the number of represented objects to the total number of real-world objects known to population completeness.
\end{itemize}
Completeness assessment particularly benefits from intertwined human-machine collaboration, because deficiencies of automated or manual completeness assessment methods can be mitigated by the complementary strategy.
  
\subsection{Level 3: Representation}
At this level, the KG should fulfill the following requirements:
\begin{itemize}
\item \textbf {Reusability (Essential)}:
\label{subsec:reusability}
Wilkinson et al.~\cite{Wilkinson2016} state that for data to be reusable, it should be:
\begin{itemize}
\item its meta (data) is thoroughly specified with a variety of precise and important properties.
\item its (meta) data are published with a clear and accessible utilization agreement.
\item its (meta) data are coupled with comprehensive provenance.
\item its (meta) data conform to community-relevant domain guidelines.
\end{itemize}
Moreover, Hogan et al.~\cite{hogan_knowledge_2022} state that reusability is the ability of a dataset to be coupled with other datasets. 
We have marked reusability as ``essential” because it is one of the main FAIR principles.
In addition, reusability emphasizes machine-actionability so that the data consumers can use the KG without any or little human intervention.
\item \textbf {Conciseness (Essential)}: 
According to Pablo et al.~\cite{Mendes2012SieveLD}, a KG is concise when there are no non-essential schema and data elements in the given KG.
Hidalgo Delgado et al.~\cite{Hidalgo-Delgado_quality_2021} differentiate between:
\begin{itemize}
\item \textbf{Intentional conciseness (schema level)}: meaning the absence of redundant schema components (properties, classes, shapes, etc.) in the KG.
\item \textbf{Extensional conciseness (data level)}: meaning the absence of duplicated entities and relations in the KG.
\end{itemize}
Batini et al.\cite{batini_methodologies_2009} refer to conciseness as ``uniqueness." They define it as the degree to which data is devoid of duplicates in areas of breadth, depth, and scope.
\cite{Frber2011SwiqaA} considers OWL helpful to avoid data redundancy on the semantic web. OWL can be used to flag synonymous identifiers with \textit{owl:sameAs}.
Batini and Scannapieco~\cite{Batinidata2006} see that when more than one state of the information system matches a state of the real-world system, data values are consistent. 
We mark conciseness as ``essential" because the more prominent the KG's conciseness, the more readable and understandable the KG is.
\item \textbf {Data representation (Useful)}:
\label{subsec:dataRepresentation}
Batini and Scannapieco~\cite{Batinidata2006} assert that the KG should present the data in a suitable language and unit, with explicit data definitions. They differentiate between different levels of data representation:
\begin{itemize}
    \item \textbf{Concise representation}: the KG presents data concisely without being encumbered.
    \item \textbf{Representational consistency}: Data is always displayed in the same format and is compatible with the older data.
\end{itemize}
\end{itemize}
\subsection{Level 4: Stability}
The KG should be available using open W3C standards\footnote{\url{https://www.w3.org/standards/semanticweb/query}} (e.g., RDF or SPARQL). In particular, the following requirements should be fulfilled:
\begin{itemize}
  \item \textbf{Trackability (Essential)}: \cite{bonatti_robust_2011,dividino_querying_2009,K-Automorphism} agree that keeping track of where the data originates from is of paramount importance when working with KGs. So it is ``essential” to verify the data sources in KGs to have a better reputation for accuracy, truthfulness, and fairness in the KG.
  \item \textbf {Identifier stability (Important)}: 
  The WC3\footnote{\url{https://www.w3.org/TR/cooluris/}} recommends using URIs as a distinctive identifier for real-world objects. In addition, Hogan et al.~\cite{hogan_knowledge_2022} state that when the KG should be augmented with external data sources, it is necessary to use globally unique IDs to avoid name conflicts and use external identity connections to distinguish a node from an external source. They provide examples of stable identifiers like Digital Object Identifiers (DOIs) for papers, ORCID IDs for authors, International Standard Book Numbers (ISBNs) for books, Alpha-2 codes for counties, and other persistent identifier (PID) schemes.
  \item \textbf {Queryability (Important)}: 
  The KG should furnish a SPARQL, GraphQL and/or API endpoint to make it straightforward for the data consumers to retrieve data from the KG.
  According to Hidalgo Delgado et al.,~\cite{Hidalgo-Delgado_quality_2021} queryability necessitates data availability via SPARQL endpoints and RDF dumps.
  SPARQL endpoints likewise streamline federated search across several data sources, enhancing and enriching data accessibility.
\end{itemize}
\subsection{Level 5: Linkability}
At this level, the KG should achieve the following prerequisites:
\begin{itemize}
 \item \textbf {Dereferencability (Useful)}: 
 Hidalgo-Delgado et al.~\cite{Hidalgo-Delgado_quality_2021} state that we can dereference resources based on URIs. We can correspondingly address the URIs using HTTP calls to return appropriate and authentic data. The resource dereferencing is complete if the HTTP call yields an RDF document and the HTTP status code is 200.
 \item \textbf {Linkability (Useful)}: According to Hogan et al.,~\cite{hogan_knowledge_2022}, linkability is the degree to which data set instances are connected. This measure is again a good example, where human-machine collaboration is beneficial. Automated linking techniques can be applied by machines using human input for training, validation and coherence assessment.
\end{itemize}

\section{Application of KGMM to a Large-Scale Use Case}
\label{sec:use-case}

The Open Research Knowledge Graph (ORKG)~\cite{jaradeh_open_2019} represents a perfect testbed for implementation and experimentation with the KGMM.
The ORKG comprises machine extracted and human curated semantic descriptions of research contributions from more than 10.000 scientific articles in 500 scientific fields~ \cite{karras2021researcher}.

A key element of the ORKG are tabular research contribution comparisons give an overview on the state-of-the-art for a certain research problem~\cite{oelen_generate_2020}. 
We implemented the KGMM for the ORKG in particular also for evaluating the maturity of comparisons.
In this section, we first clarify how the ORKG satisfies the KGMM's requirements and then illustrate in detail how we implemented the KGMM for ORKG comparisons.
\begin{figure}
  \centering
  \includegraphics[width=\linewidth]{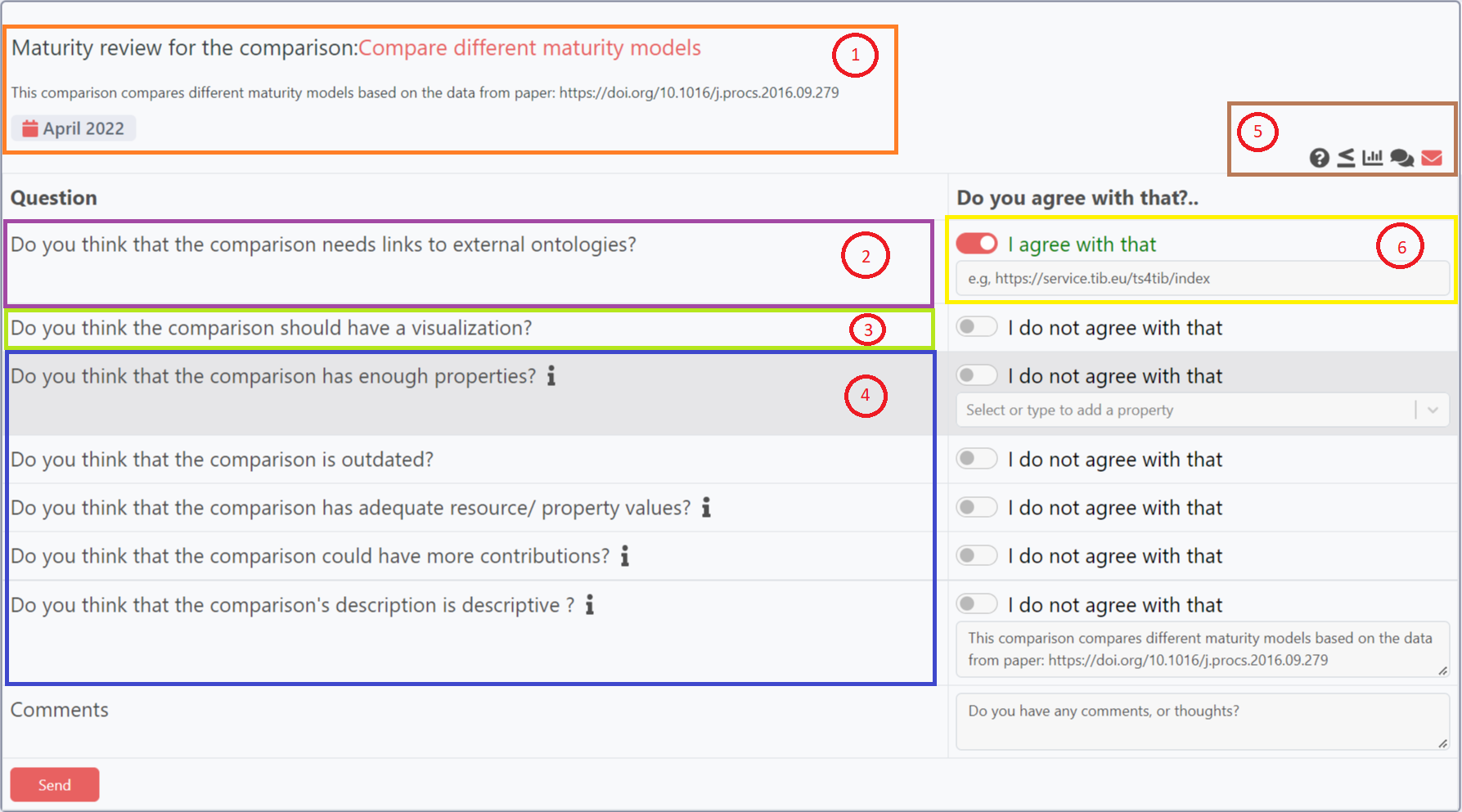}
  \caption{\textit{Comparison peer review form for feedback from crowd members.} Box one (orange) contains the comparison's metadata. Box two (purple) contains a question asking the reviewers about the linkability (KGMM level 5). The question in box three (lime) asks the reviewers about the data representation (KGMM level 3). Questions in box four (blue) ask the reviewers about the KG completeness (KGMM level 2). The icons in box five (brown) enable the user to find all the needed reports or send the review to another expert. Box six (yellow) let the reviewers add links to external ontologies, if they can recommend any.}
  \label{fig:comparisonReview}
\end{figure}

\subsection{ORKG Maturity Levels}
The ORKG comparison has five maturity levels, identical to the KGMM model. 
   We have illustrated a detailed table that demonstrates how the ORKG implemented the various KGMM maturity levels, quality dimensions, metrics, and measures available on Zenodo. \cite{hussein_hassan_2022_6732786}
\subsection{Application of KGMM to ORKG Comparisons}

One crucial type of content in the ORKG are comparisons of research contributions addressing a specific research question. 
We now discuss how the KGMM was specifically applied to this type of content.

\begin{figure}
\centering
\includegraphics[trim={0 .2cm 0 1.1cm},clip,width=\linewidth]{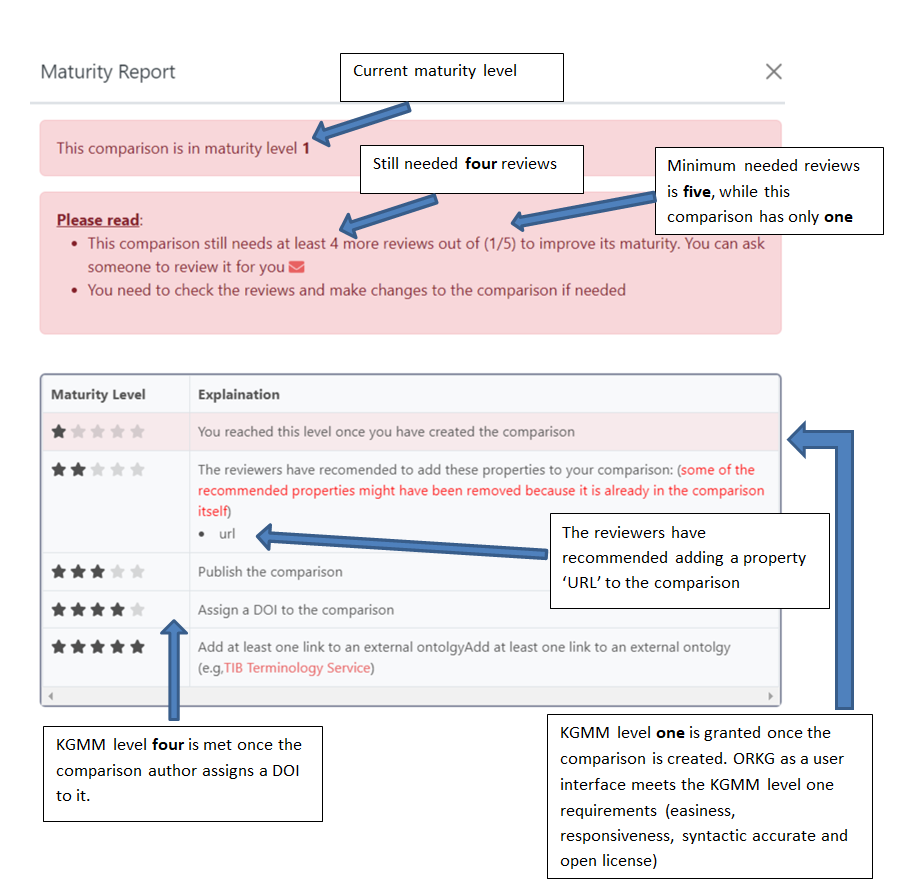}
\caption{\textit{Full report about the current maturity level, and the number of added reviews.} The report aims at answering the following questions:
(1) What is the current maturity level?
(2) What are the properties, that the reviewers have recommended?
(3) How many reviews in total are needed for a comparison?
(4) How many reviewers have reviewed the comparison so far?
}
\label{fig:maturityReport}
\end{figure}

\vspace{-.2cm}
\paragraph{The reviews.} Once a comparison is published in the ORKG, peer community members (other than the original author) can add reviews for this comparison.
The comparison author can invite others to review the comparison by sending an email with the review URL.
The comparison review is publicly accessible for all users resembling an open review process. 
To ensure data quality and provenance tracking, only registered users can add reviews.

We followed the design approach proposed by Cao et al.~\cite{qi_knowledge_2021} regarding to crowdsourcing task design.
Implicit crowdsourcing tasks excel better in terms of cost and outcomes.
The implicit crowdsourcing task implies that the crowd members are unaware of the existence of the task since it is part of the genuine user interaction with the application.
Additionally, they suggest designing binary choice (true or false) questions for more acceptable results.
As shown in \autoref{fig:comparisonReview}, our implemented review is a relatively straightforward, implicit agree/disagree task, where the reviewers inform us of their thought on the constructed comparison.

\vspace{-.2cm}
\paragraph{Determination of the minimum review number by the community.} We do not know how many reviews would be adequate to review a given comparison. A reasonable solution here is to allow the community to determine that by themselves. We enable the crowd members to decide what the minimum of required reviews should be for a comparison in their field.

\vspace{-.2cm}
\paragraph{The maturity report.} The report illustrates the maturity level of a particular comparison. Moreover, the report answers the question ``Why is a particular comparison at a certain maturity level?" The report helps the comparison author to comprehend what should be done to advance the maturity level by providing concrete steps. Also, the report is publicly available to the community, so others can help to improve the maturity of the comparison (cf. \autoref{fig:maturityReport}).

\vspace{-.2cm}
\paragraph{The feedback report.} The report reveals how the reviewers evaluate a given comparison. It is an aggregative, publicly available report of the reviewer’s feedback for a given comparison. The report draws the comparison author’s attention to what the reviewers think about his comparison. On the other hand, the report also assists the community in drawing an image of the others' thoughts.

\section{Conclusions and Future Work}
\label{sec:conclusions}

This work is a first of a larger research and development agenda.
We want to lift knowledge graphs to a novel level, by more 1) systematically intertwining human and machine intelligence and 2) developing means for managing more complex information assets in KGs then mere entity descriptions.
For both of these aspects, a maturity model for gradually improving the representations in the KG is of utmost importance.
With KGMM we presented a first version of a graded maturity model for knowledge graph applications aiming to intertwine human-machine collaboration.
The proposed model consists of 5 maturity levels, comprising 20 quality measures. 
It has a strong foundation in existing quality assessment frameworks, including the FAIR data guidelines, LOD stars, and a set of additional work related to linked data quality. 
The maturity model is complemented with dimensions specifically related to human-machine curated knowledge graphs.

We plan also to do either a qualitative or empirical evaluation. In that sense, we have to wait for one to one and a half years until we have adequate peer-reviews for the KGMM by our users. Afterward, we can have an extended version of the paper as a journal paper describing a detailed evaluation of the KGMM.
%
%
%
\paragraph*{Supplemental Material Statement:} Source code for Section \ref{sec:use-case} is available on Github.\footnote{\url{https://gitlab.com/TIBHannover/orkg/orkg-frontend/-/merge_requests/883}}
\bibliographystyle{splncs04}
\bibliography{refs}
\end{document}